\documentclass[11pt]{article}
\usepackage{amsmath,amssymb}
\usepackage{float}
\usepackage{caption2}
\usepackage{graphicx}
\def\be{\begin{equation}}
\def\ee{\end{equation}}
\def\ba{\begin{eqnarray}}
\def\ea{\end{eqnarray}}

\begin{document}
\title{\textbf{\large{Static Solutions of Einstein's Equations with
 Spherical Symmetry }}}
\vspace{1cm}
\author{IftikharAhmad\thanks{e-mail:dr.iftikhar@uog.edu.pk}, Maqsoom
Fatima and Najam-ul-Basat.
 \\
\textit{Institute of Physics and Mathematical Sciences}, \\
\textit{Department
of Mathematics, University of Gujrat}\\
\textit{ Pakistan.}}

\date{}
\maketitle
\begin{abstract}
The Schwarzschild solution is a complete solution of Einstein's
field equations for a static spherically symmetric field. The
Einstein's field equations solutions appear in the literature, but
in different ways corresponding to different definitions of the
radial coordinate. We attempt to compare them to the solutions with
nonvanishing energy density and pressure. We also calculate some
special cases with changes in spherical symmetry.

\end{abstract}
\maketitle

\section{Introduction}
Even though the Einstein field equations are highly nonlinear
partial differential equations, there are numerous exact solutions
to them. In addition to the exact solution, there are also non-exact
solutions to these equations which describe certain physical
systems. Exact solutions and their physical interpretations are
sometimes even harder. Probably the simplest of all exact solutions
to the Einstein field equations is that of Schwarzschild.
Furthermore, the static solutions of Einstein's equations with
spherical symmetry (the exterior and interior Schwarzschild
solutions) are already much popular in general relativity, solutions
with spherical symmetry in vacuum are much less recognizable.
Together with conventional isometries, which have been more widely
studied, they help to provide a deeper insight into the spacetime
geometry. They assist in the generation of exact solutions,
sometimes new solutions of the Einstein field equations which may be
utilized to model relativistic, astrophysical and cosmological
phenomena. Stephani et al.(\cite{SKMH})has emphasized the role of
symmetries in classifying and categorizing exact solutions.\\
Symmetries are used as one of the principal classification schemes
in their catalogue of known solutions. In this paper we construct
the vacuum solutions of this sort, and we set up the differential
equations for nonvacuum solutions. Indeed, to some extent more
general vacuum solutions, possibly breaking the translational
invariance, were found in the early $20th$ century by Weyl and
Levi-Civita (\cite{WL} \cite{RF}), and their analogs breaking the
static condition were studied by Rosen and Marder \cite{BJ} in
mid-century. Afterward, much attention was given to ``cosmic
strings'' thin cylinders (usually filled with a non-Abelian gauge
field) surrounded by vacuum (e.g. [{\cite{BB}}-{\cite{Ei}}]).

In this paper the metric obtained describes the solution in vacuum
due to spherically symmetric distribution of matter. The field is
static and can be produced by spherically  symmetric motion. The
requirement of spherical symmetry alone is sufficient to yield the
static nature of our solution. Moreover, the metric tensor tends to
approach the Minkowskian flat spacetime metric tensor, and also the
well known cosmic string solutions are locally flat (;regular
Minkowskian spacetime minus a wedge described by a deficit angle).
It is not widely appreciated that the static, translationally
invariant cases of the older solutions (
\cite{WL},\cite{RF},\cite{BJ} ) are
not all of that type.\\
We write the most general expression for a spacetime metric with
static and spherical symmetry and solve the Einstein field equations
for the components of the metric tensor. We carefully remove all
redundant solutions corresponding to the freedom to rescale the
coordinates, thus the principal result is the
general solution.\\
Like the exterior Schwarzschild solution (when it is not treated as
a black hole) one expects these spherical solutions to be physically
related only over some subinterval of the radial axis. The easiest
to find are the cosmic string solutions of Gott and others (
\cite{BB} \cite{BB1} ), which have the locally flat cone solution on
the outside and a constant energy density $\rho$ inside, with
pressure $p_{\phi} =-\rho$ along the axis and vanishing pressures
$p_{r} = p_{\theta} = 0$ in the perpendicular plane. Although
natural from the point of view of gauge theory \cite{Ei}, such an
equation of state would be surprising for normal matter.
\section{Static Solutions of Einstein's Equations with Spherical Symmetry}
\subsection{A Vacuum Spherical Solution of Equations}
A general expression for writing a metric exhibiting spherical
symmetry, we require that it must have axial symmetry and thus the
metric components must be independent of $\phi$. As we are also
examining only the static phase of universe, so that the metric
components must be independent of cosmic time $t$, leaving any
unknown functions to be functions of the radial variable $r$ only.
In analogy to the standard treatment of spherical symmetry, we
define $r$ such that the coefficient of $d\phi^2$ is equal to
$r^2sin^2\theta$. Thus the metric can be written as (see Ref.
\cite{Y})

\begin{eqnarray}\label{1a}
  ds^2 &=& -e^{2\Phi}dt^2+e^{2\Lambda}dr^2
   +r^2d\theta^2+r^2sin^2\theta
 d\phi^2.
\end{eqnarray}
Where $\Phi$, $\Lambda$ are the unknown functions of $r$ for which
we should like to solve. By writing our unknown functions in the
form of exponentials, we guarantee that our coefficients will be
positive as we would like them to be, and also mirror the standard
textbook treatment of the spherically symmetric metric. The form in
which we have written the metric does not restrict the range of
$\theta$ to be from $0$ to $\pi$; instead it runs from $0$ to some
angle $\theta_*$.\\\\
Using the standard known expressions for the Christoffel symbols
$\Gamma^\rho_{\mu\nu}$, Riemann curvature tensor
$R^\rho_{\lambda\mu\nu}$, Ricci tensor $R_{\mu\nu}$, Einstein tensor
$G_{\mu\nu}$ and Stress tensor $T^{\mu\nu}$ associated with a given
metric \cite{Y}, all of the components of these objects can be
calculated for this static, spherically symmetric metric. The
results for this are presented
below.\\
\newpage
\textbf{ Nonzero Christoffel Symbols:}

\begin{eqnarray*}
  \Gamma^t{_t}{_r} &=& \Gamma^t{_r}{_t}=\Phi' \\
  \Gamma^r{_t}{_t} &=& \Phi'e^{2(\Phi-\Lambda)} \\
 \Gamma^r{_r}{_r} &=& \Lambda' \\
 \Gamma^r{_\theta}{_\theta} &=& -re^{-2\Lambda} \\
 \Gamma^r{_\phi}{_\phi} &=& -e^{-2\Lambda}rsin^2\theta \\
 \Gamma^\phi{_r}{_\phi} &=& \Gamma^\phi{_\phi}{_r}={1\over r}\\
  \Gamma^\theta{_r}{_\theta}&=& \Gamma^\theta{_\theta}{_r}={1\over r}\\
  \Gamma^\theta{_\phi}{_\phi}&=&- sin\theta cos\theta\\
  \Gamma^\phi{_\phi}{_\theta}&=& cot\theta\\
\end{eqnarray*}
\textbf{Nonzero Riemann Curvature Tensor Components:}
\begin{eqnarray*}
  R^t{_\theta}{_\theta}{_t} &=& r\Phi'e^{-2\Lambda} \\
  R^r{_\theta}{_\theta}{_r} &=& -r\Lambda'e^{-2\Lambda} \\
  R^r{_\phi}{_\phi}{_r} &=& -\Lambda're^{-2\Lambda}sin^2\theta \\
  R^r{_t}{_t}{_r} &=& -(\Phi''+{\Phi'}^2-\Phi'\Lambda')e^{2(\Phi-\Lambda)} \\
  R^\phi{_t}{_t}{_\phi} &=& -{1\over r}\Phi'e^{2(\Phi-\Lambda)}\\
  R^\phi{_\theta}{_\theta}{_\phi}&= & e^{-2\Lambda}-1\\
\end{eqnarray*}
\textbf{Nonzero Ricci Tensor Components:}
\begin{eqnarray*}
  R{_t}{_t} &=& (\Phi''+{\Phi'}^2-\Phi'\Lambda'+{2\over r}\Phi')e^{2(\Phi-\Lambda)} \\
  R{_r}{_r} &=& -\Phi''+\Lambda'\Phi'-{\Phi'}^2+{2\over r}\Lambda' \\
  R{_\theta}{_\theta} &=& (-r\Phi'-1+r\Lambda')e^{-2\Lambda}+1\\
R{_\phi}{_\phi}&=&[(-r\Phi'-1+r\Lambda')e^{-2\Lambda}+1]sin^2\theta\\
\end{eqnarray*}
Where primes correspond to differentiation with respect to $r$,
e.g., $\Phi'={d\Phi\over dr}$ and $\Lambda'={d\Lambda\over dr}$.\\
We should like to solve the Einstein field equations for the vacuum
solution, which corresponds to $G{_\alpha}{_\beta}=0$.
However, that is sufficient to calculate the solutions for
$R{_\alpha}{_\beta}=0$. We begin with the standard
 definition of the Einstein tensor,
 $G{_\alpha}{_\beta}=R{_\alpha}{_\beta}-{1\over
 2}Rg{_\alpha}{_\beta}$.\\
 Using this expression for mixed tensor, someone can calculate the trace of the
 Einstein tensor $G_{\mu\nu}$ with $\delta^\mu_\mu=4$;~~~~$$G^\mu{_\mu}=R^\mu{_\mu}-{1\over
 2}R \delta^\mu_{\mu}=R-2R=-R.$$ And we obtain the following relation
 between the Ricci and Einstein tensors:
 $$R{_\alpha}{_\beta}=G{_\alpha}{_\beta}-{1\over
 2}Gg{_\alpha}{_\beta}.$$ Thus we see that if $R{_\alpha}{_\beta}=0$
 then $G{_\alpha}{_\beta}=0$, and conversely, if
 $G{_\alpha}{_\beta}=0$  then $R{_\alpha}{_\beta}=0$. We concluded that the
 solutions to $R{_\alpha}{_\beta}=0$ are also the solutions to the
 vacuum Einstein field equations, $G{_\alpha}{_\beta}=0$.\\\\
 ~~~~~~By equating the nontrivial components of the Ricci tensor
 with zero, we find a set of three ordinary differential equations
 for $\Phi$ and $\Lambda$. We further see that the
 exponential function is never equal to zero, so the differential
 equations reduce to
\ba \Phi''+{\Phi'}^2-\Phi'\Lambda'+{2\over r}\Phi'=0 \label{6a}\ea
 \ba -\Phi''+\Lambda'\Phi'-{\Phi'}^2+{2\over r}\Lambda'=0 \label{7a}\ea
 \ba (-r\Phi'-1+r\Lambda')e^{-2\Lambda}+1=0 \label{8a}\ea

\begin{figure}[t]
\begin{center}
\includegraphics[width=7cm]{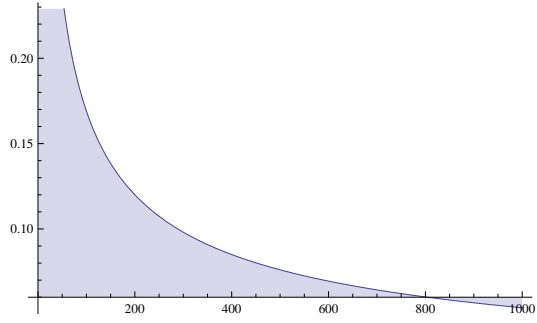}
\caption{ The evolution of $\Phi$, for $r_G=2.99~km$.
 }\label{xx}
\end{center}
\end{figure}

Adding Eq. (\ref{6a}) and Eq. (\ref{7a}), to get $\Phi'=-\Lambda'$
then substituting this value in Eq. (\ref{8a}). Thus this system can
be reduced to \ba r e^{2\Phi}=r\pm a,\label{AA}\ea where $a$ is
constat.

\begin{figure}[t]
\begin{center}
\includegraphics[width=7cm]{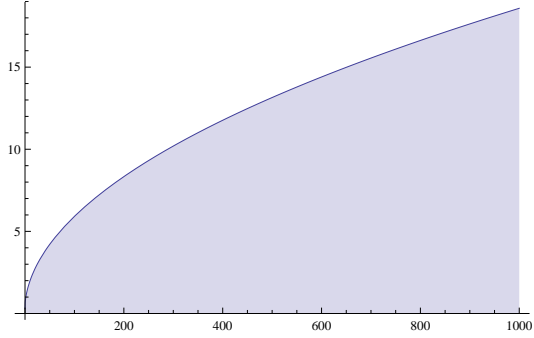}
\caption{ The evolution of $\Lambda$, with different $r_G=2.99~km$.
 }\label{xx}
\end{center}
\end{figure}


\section{Some Special Cases}
Here we examine the existence of some special points by considering
it's geometry and observe their importance in different forms of
spherical metric.\\

$\bullet$ \textbf{Case I:} When $r\rightarrow \infty$, very large
enough, in the case of accelerating universe, $e^{2\Phi}=1$,
~~~~$e^{2\Lambda}=1$, then we obtain the metric

\ba ds^2=-dt^{2}+dr^2+r^{2}[d\theta^2+sin^2\theta
d\phi^2]\label{18a},\ea this describes a Minkowskian (spacetime)line
element outside matter distribution.\\$\bullet$ \textbf{Case II:}
When $a=r_G$ gravitational radius of the body. Thus the metric is
reduced to
$$ds^2=-(1-{{r_G}\over{r}})dt^{2}+(1-{{r_G}\over{r}})^{-1}dr^2+r^{2}[d\theta^2+sin^2\theta
d\phi^2],$$ is called Schwarzschild line element.\\$\bullet$
\textbf{Case III:} When $a=r=r_G$, then $g_{tt}=0$ and
$g_{rr}=\infty$. Thus it should be noted that Schwarzschild metric
become singular at that point. However, for most of the observable
bodied in the universe the gravitational radius lies well inside
them. For example in the case of Sun the value of $r_G= 2.9 km$ and
for our Earth its value is $0.88 cm$. \\$\bullet$ \textbf{Case IV:}
At $a=r$, we have $e^{2\Phi}=1+{{a}\over{r}}$, then metric reduces
to the form \ba
ds^2=-2dt^2+{{1}\over{2}}dr^2+r^2(d\theta^2+sin^2\theta
d\phi^2)\label{20a}\ea this describes a very special metric of the
Universe and its properties are not yet discussed in this paper.
\section{Solutions of the Einstein Equation with Sources}
To find some spherical space times that are not singular along the
central axis. We need to solve Einstein equations, where stress
energy tensor $T$ has nonzero components, first we have to find some
basic quantities and tensors encountered in general relativity. For
this we find the Ricci scalar, $R$, the Einstein tensor,
$G_{\mu\nu}$, and the stress energy tensor, $T_{\mu\nu}$, for the
spherically
symmetric metric given in (\ref{1a})\\
\textbf{Ricci scalar:} \ba
R=e^{-2\Lambda}\left[-2\Phi''-2{\Phi'}^{2}+2{\Lambda'}{\Phi'}-4\frac{\Phi'}{r}
+4\frac{\Lambda'}{r}-\frac{2}{r^{2}}\right]+{{2}\over{r^2}}\label{21a}\ea
 \textbf{Nonzero Einstein Tensor Components are:}
\begin{eqnarray*}
  G_{tt} &=&
e^{2(\Phi-\Lambda)}(\frac{2\Lambda'}{r}-\frac{1}{r^{2}})+{{1}\over{r^2}}e^{2\Phi}  \\
  G_{rr} &=& \frac{2\Phi'}{r}+\frac{1}{r^{2}}-{{1}\over{r^2}}e^{2\Lambda} \\
  G_{\theta\theta} &=& e^{-2\Lambda}{r^{2}}\left[\Phi''+{\Phi'}^{2}-{\Lambda'}{\Phi'}+\frac{\Phi'}{r}
-\frac{\Lambda'}{r}\right]
\end{eqnarray*}
\ba
G_{\phi\phi}=e^{-2\Lambda}{r^{2}}sin^{2}{\theta}\left[\Phi''+{\Phi'}^{2}-{\Lambda'}{\Phi'}+\frac{\Phi'}{r}
-\frac{\Lambda'}{r}\right].\label{22a}\ea

The stress energy tensor components are defined by
$T_{{r}}^{r}=p_{r}$  and $T_{rr}=p_{r}e^{2\Lambda}$, similarly the
remaining pressure components are defined, similarly also we have
$T_{{t}}^{t}=-\rho$\\
\textbf{Nonzero Stress Tensor Components are:}
$$T_{tt}=\rho{ e^{2\Phi}}$$
$$T_{rr}=p_{r} {e^{2\Lambda}}$$
$$T_{\theta\theta}=p_{\theta}{r^{2}}$$
\ba T_{\phi\phi}=p_{\phi}{r^{2}}sin^{2}{\theta}\label{23a}\ea
\\In General Relativity (GR), the symmetric stress-energy tensor
acts as the source of spacetime curvature. While dealing with the
curved spacetime due to the existence of matter, the Riemann tensor
plays a vital role as seen in GR. One very important equation in
this subject is the Einstein field equation, a tensorial equation
which takes the form $G_{\mu\nu}=8\pi T_{\mu\nu}$,  $G_{\mu\nu}$ is
an Einstein's tensor which is symmetric and vanishes when spacetime
is flat, $T_{\mu\nu}$ is the so-called energy-momentum tensor which
can be thought of as a source for the gravitational field. It is a
divergenceless tensor due to the conservation of energy, namely
$\nabla^{\mu}T_{{\mu\nu}}=0$. The proportionality constant is $8\pi$
since we use the natural units, otherwise it would be $\frac{8\pi
G}{c^{4}}$.\\\\ Mathematically, the Einstein's tensor is given by
$G_{\mu\nu}=R_{\mu\nu}-\frac{1}{2}g_{\mu\nu}R$,  where Ricci tensor
is a contraction of the Riemann tensor $(R_{\mu\nu} =
R_{{\gamma}}^{\mu\gamma\nu})$, and R is a curvature scalar obtained
from the Ricci tensor, hence also called Ricci scalar. The full form
of the Einstein equation has an extra term owing to the Cosmological
constant $(\Lambda^*)$, which has been found recently to be an
extremely tiny number but non-zero. It reads
$G_{\mu\nu}-{\Lambda^*}g_{\mu\nu}=8\pi T_{\mu\nu}$ The significance
of the cosmological constant is involved mostly in the context of
cosmology in which one studies the fate of the universe (for example
see Ref. {\cite{NP}}), from $T^{\mu\nu}_{;\nu}=0$ we have \ba
0=\frac{\partial p_{r}}{\partial r}+p_{r}(\Phi'+\frac{2}{r})+\rho
\Phi'-\frac{p_{\theta}}{r}-\frac{p_{\phi}}{r}\label{24a}\ea
\ba4\pi(\rho+p_{r}+p_{\theta}+p_{\phi})e^{2\Lambda}=\Phi''+{\Phi'}^{2}-{\Lambda'}{\Phi'}+\frac{2\Phi'}{r}
\label{25a}\ea \ba
4\pi(\rho+p_{r}-p_{\theta}-p_{\phi})e^{2\Lambda}=-\Phi''-{\Phi'}^{2}+{\Lambda'}{\Phi'}+\frac{2\Lambda'}{r}\label{26a}\ea
\ba
4\pi(\rho-p_{r}+p_{\theta}-p_{\phi})e^{2\Lambda}=\frac{\Lambda'}{r}-\frac{\Phi'}{r}
-\frac{1}{r^{2}}+{{1}\over{r^2}}e^{2\Lambda}\label{27a}\ea \ba
4\pi(\rho-p_{r}-p_{\theta}+p_{\phi})e^{2\Lambda}=\frac{\Lambda'}{r}-\frac{\Phi'}{r}
-\frac{1}{r^{2}}+{{1}\over{r^2}}e^{2\Lambda}\label{28a}\ea

Now adding equation (\ref{25a}) and (\ref{26a}) then subtracting
equation (\ref{28a}). This gives \ba
4\pi(\rho+3p_{r}+p_{\theta}-p_{\phi})e^{2\Lambda}=3\frac{\Phi'}{r}
+\frac{\Lambda'}{r}+\frac{1}{r^{2}}-{{1}\over{r^2}}e^{2\Lambda}\label{29a}\ea
Now we add and subtract equation (\ref{27a}) in equation (\ref{29a})
and get simultaneously results, \ba
4\pi(2\rho+2p_{r}+2p_{\theta}-2p_{\phi})e^{2\Lambda}=2\frac{\Phi'}{r}
+2\frac{\Lambda'}{r}\label{30a}\ea
 \ba
4\pi(4p_{r})e^{2\Lambda}=4\frac{\Phi'}{r}
+\frac{2}{r^{2}}-{{2}\over{r^2}}e^{2\Lambda}\label{31a}\ea

Here we have a system of differential equations namely Eqs.
(\ref{24a}), (\ref{25a}), (\ref{27a}) and (\ref{28a}) can be solved
for $p_{r}\;\;, p_{\theta}\;\;, p_{\phi} \;\;,\Phi\;\;and\;\;
\Lambda$. Further Eq. (\ref{31a}) which contains only first order
derivative of the unknown function, posses an additional constraint.
Thus the above system of five equations is second order in $\Phi$
and first order in $\Lambda$ and $p_{r}$.\\ By taking derivative of
equation (\ref{31a}) with respect to $r$ and putting values of
equations (\ref{24a}), (\ref{25a}), (\ref{28a}) and (\ref{30a}) to
substitute for $p'_{r}$, $\Lambda'$ and $\Phi''$ resulting a
relation which reduces $0=0$ thus equation (\ref{31a}) would must
hold for all $r$ if it holds at any value of $r$ .

\section{A Solution with $\rho=-p_{\phi}\;\;, p_{r}=p_{\theta}=0$}
The generic solution of these differential equations for arbitrary
values of our unknown $\rho,\;\;p_{r},\;\;p_{\theta}\;\; and\;\;
p_{\phi}$ is much more difficult and out of scope of this paper. For
simplicity of finding the solution of equation, one can take the
value $\rho=-p_{\phi}$ and the remaining pressure components are
zero. Thus from equations (\ref{24a}), (\ref{25a}), (\ref{27a}),
(\ref{28a}), (\ref{30a}) and (\ref{31a}) respectively, someone can
get

\ba 0=\rho\left(\Phi'+\frac{1}{r}\right)\label{32a}\ea \ba
0=\Phi''+{\Phi'}^{2}-{\Lambda'}{\Phi'}+\frac{2\Phi'}{r}
\label{33a}\ea \ba
4\pi(2\rho)e^{2\Lambda}=\frac{\Lambda'}{r}-\frac{\Phi'}{r}
-\frac{1}{r^{2}}+{{1}\over{r^2}}e^{2\Lambda}\label{34a}\ea \ba
0=\frac{\Lambda'}{r}-\frac{\Phi'}{r}-\frac{1}{r^{2}}+{{1}\over{r^2}}e^{2\Lambda}\label{35a}\ea
\ba 4\pi(2\rho)e^{2\Lambda}=\frac{\Lambda'}{r}+\frac{\Phi'}{r}
\label{36a}\ea and  \ba
0=2\frac{\Phi'}{r}+\frac{1}{r^{2}}-{{1}\over{r^2}}e^{\Lambda}\label{36a}\ea

From Eq. (\ref{32a}), we get $\Phi=\ln[{1/r}]$ and with the help of
Eq. (\ref{32a}), we obtain from Eq. (\ref{34a}) $\Lambda'={{(8\pi
\rho r^2-1)}\over{r}} e^{2\Lambda},$ thus the metric of the solution
is \ba ds^2=-r^{-2}dt^2+e^{2\Lambda}dr^2+r^2d\theta^2+r^2sin^2\theta
d\phi^2\label{37a}.\ea \\ If we take the value of $\rho=1/{8\pi
r^2_0}$, which is well known result in literature, then our final
solution becomes \ba
ds^2=-r^{-2}dt^2+[r^2_0/(r^2_0ln{r^2}-r^2)]dr^2+r^2d\theta^2+r^2sin^2\theta
d\phi^2\label{38a}.\ea
\begin{figure}[t]
\begin{center}
\includegraphics[width=8cm]{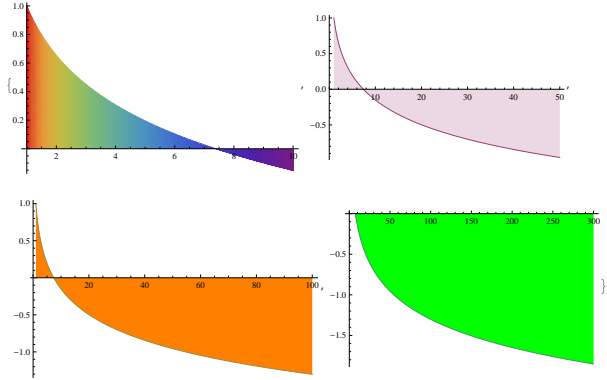}
\caption{ The evolution of $\Phi$, with different panels,
representation of Universe.
 }\label{xx}
\end{center}
\end{figure}

\section{Discussion}
Spherical symmetry in general relativity turns out to be similar to
Cylindrical symmetry in many of its behavior but very different in
other actions.. We have presented analytical solution with spherical
symmetric approach with more conventional interior sources and more
general exterior geometry. However the construction of a variety of
solutions presented here illustrates several informative points
about metric of the Universe. On the basis of information we would
like to decide the ultimate fate of the Universe. \\Furthermore, the
choice of gauge (coordinate system) is always a major issue in
relativity; the same space-time can look quite different in
different gauges, and how (and whether) to choose a standard gauge
or ``normal form" for a given problem is not always understandable.
There are many acceptable ways to fix the gauge, and we have taken
pains to describe them all and how they are related. Even after a
definition of radial coordinate has been selected, further steps to
a normal form can be taken by linear rescaling of the coordinates.
The structure of the Einstein equation system is nontrivial. There
is one more equation than one might naively expect. The extra
equation serves as a constraint on the data. For the spherical
vacuum solutions this constraint is a simple algebraic relation
among the parameters.\\ Finally, we observed some surprising
ambiguities of interpretation. For the evolution of $\Phi$ (see
figure 1) an increasing radius, feasible region is disappear from
actual frame for fixing the value of $r_G=2.99~km$, but in the case
of evolution of $\Lambda$ (see figure 2 ) shows increasing behavior
due to increase of $r$ in the given metric. However, we observed
that when $r$ is large enough, then actual region in upper half is
decreasing, ratio of two reign is decreasing. It means that ratio is
directly proportional to $r$, therefore, we see from (figure 3) that
in below left panel upper region will be completely disappear, which
is the consistency of our theoretical results.\\
Furthermore, numerical solution with different parameters of the
metric still need special attention for future work. In future work,
we intend to investigate the compatibility of the conformal
spherical symmetry with homogeneity, and also with the kinematical
quantities.   ~~ ~~
\section{Acknowledgment}
The author would like to thanks Dr. Muhammad Nizamuddin for providing the
facilities to carry out the research work and we thank Yun-Song Piao for useful suggestion.

\end{document}